\documentclass[aps,pra,a4paper,twocolumn,longbibliography, amsmath,amssymb,groupedaddress]{revtex4-2}
\usepackage[colorlinks=false, allcolors=blue]{hyperref}
\usepackage{graphicx}
\usepackage{xcolor}
\usepackage[nonumberlist, acronym, toc]{glossaries}

\usepackage[margin = 1in]{geometry}

\begin{document}

\glsdisablehyper

\newacronym{naqpu}{NA-QPU}{Neutral Atom - Quantum Processing Unit}
\newacronym{qpu}{QPU}{Quantum Processing Unit}
\newacronym{ftqc}{FTQC}{Falt-Tolerant Quantum Computer}

\newacronym{de}{DE}{Differential Equation}

\newacronym{nn}{NN}{neural network}
\newacronym{cnn}{CNN}{convolutional neural network}
\newacronym{rnn}{RNN}{recurrent neural network}
\newacronym{lstm}{LSTM}{Long Short-Term Memory}
\newacronym{mlp}{MLP}{Multi-Layer Perceptron}
\newacronym{pinn}{PINN}{physics-informed neural network}
\newacronym{fm}{FM}{feature map}
\newacronym{ml}{ML}{machine learning}
\newacronym{gml}{GML}{graph machine learning}
\newacronym{sciml}{SciML}{scientific machine learning}
\newacronym{pi}{PI}{Physics-informed}
\newacronym{ufa}{UFA}{universal function approximator}
\newacronym{wl}{WL}{Weisfeiler-Lehman}
\newacronym{ibp}{IBP}{integration by parts}

\newacronym{qc}{QC}{Quantum Computing}
\newacronym{na}{NA}{Neutral Atoms}
\newacronym{nisq}{NISQ}{noisy intermediate scale quantum}
\newacronym{hea}{HEA}{Hardware Efficient Ansatz}

\newacronym{qsciml}{QSciML}{Quantum scientific machine learning}
\newacronym{qml}{QML}{quantum machine learning}
\newacronym{dqc}{DQC}{differentiable quantum circuits}
\newacronym{pqc}{PQC}{parameterised quantum circuit}
\newacronym{qcl}{QCL}{quantum circuit learning}
\newacronym{qnn}{QNN}{quantum neural network}
\newacronym{psr}{PSR}{Parameter Shift Rule}
\newacronym{agpsr}{aGPSR}{approximate Generalized Parameter Shift Rule}
\newacronym{qel}{QEL}{Quantum Extremal Learning}
\newacronym{qrc}{QRC}{quantum reservoir computing}

\title{Weak forms offer strong regularisations: \\ 
how to make physics-informed (quantum) machine learning more robust}
\author{Annie E. Paine}
\author{Smit Chaudhary}
\author{Antonio A. Gentile}
\affiliation{Pasqal, 7 rue L\'{e}onard de Vinci, 91300 Massy, France}

\begin{abstract}
Physics-informed (PI) methodologies have surged to become a pillar route to solve Differential Equations (DEs), sustained by the growth of machine learning methods in scientific contexts.  
The main proposition of PI is to minimise variationally a loss function, formally ensuring that a neural surrogate of the solution has the DE locally satisfied. 
The nature of such formulation encouraged the exploration of equivalent quantum algorithms, where the surrogate solution is expressed by variational quantum architectures. 
The locality of typical loss functions emphasises the DE to hold at an ensemble of points sampled in the domain, but encounters issues when generalising beyond such points, or when propagating boundary conditions. Issues which affect classical and quantum PI algorithms alike.
The quest to fill this gap in robustness and accuracy against mainstream DE solvers has led to a plethora of proposals in various directions. 
In particular, classical DE solvers have long employed the weak form - an integral based approach aiming at imposing a global condition on the solution - prioritising a good average behaviour instead of ``overfitting'' select points. 
Here, we propose and explore to combine contributions from both local and global loss functions in PI routines, to exploit the advantages and mitigate the weaknesses of both. 
We showcase this intuition in a variety of problems focusing on differentiable quantum architectures, and demonstrating in particular how orchestrating such hybrid loss formulation with domain decomposition can offer a strong advantage over local-only strategies.
\end{abstract}

\maketitle

\section{Introduction}

The development of quantum computers opens opportunities for improving efficiency across a variety of computationally hard problems \cite{shor1994algorithms, harrow2009quantum, Eisert2025}. While current quantum devices remain limited and require further development for their full potential to be realized, exploring possible use cases is essential for rapidly exploiting these devices as they mature. 
With several applications under active investigation targeting quantum advantage~\cite{babbush2025grand}, including the simulation of microscopic systems \cite{ ma2020quantum, clinton2024towards} and optimization tasks \cite{ moll2018quantum, zhou2020quantum}, solving \glspl{de} represents a particularly promising target for industrial quantum utility.

\acrlongpl{de} are fundamental to many fields, from fluid dynamics \cite{anderson2013computational} and financial modeling \cite{oksendal2013stochastic} to population dynamics \cite{holmes1994partial} and materials science \cite{braun1983differential}. However, solving many differential equations remains computationally expensive or intractable with classical methods \cite{ gear1981numerical, wanner1996solving, moin1998dns, jentzen2018overcoming, zhang2023turbulence}. This challenge pushed towards quantum approaches that might offer computational advantages for this ubiquitous problem class.

Variational methods hold the advantage of \gls{nisq} amenability \cite{huang2023near, cerezo2021variational}, motivating recent research in their applicability as \gls{de} solvers \cite{lubasch2020variational, sato2021variational, kyriienko2021solving, Gourianov2022, berger2025trainable, wu2025effhamiltonian, setty2025selfqpinn}.
In particular, \gls{dqc} use a \gls{qnn} to represent the solution of a given \gls{de} by training against an associated loss function~\cite{kyriienko2021solving}. \gls{dqc} can hence be viewed as a quantum counterpart to \gls{pinn} \cite{cuomo2022scientific}, but exploiting the exponentially large Hilbert space and expressivity of quantum systems to search for, and ultimately represent, the targeted solutions. 
However, \gls{dqc} can inherit challenges common to both variational quantum algorithms as well as physics-informed approaches, including barren plateaus \cite{larocca2025barren, mcclean2018barren}, convergence to local minima \cite{you2021exponentially, krishnapriyan2021characterizing}, and susceptibility to trivial solutions \cite{leiteritz2021avoid, Chuang2022pinnfrustration}.

To date, implementations of \gls{dqc} for \glspl{de} have predominantly employed a \textit{collocation}-based loss function, in which the residual of the \gls{de} is minimized at a discrete set of points in the domain \cite{kyriienko2021solving}. While this local approach enforces precise satisfaction of the \gls{de} at the chosen points, it suffers from several limitations. Its local nature can impede the propagation of boundary information throughout the domain, increasing the risk of convergence to trivial solutions that satisfy the loss function but elude the true solution. Moreover, collocation-only loss functions require additional terms when employing techniques such as domain decomposition, complicating the training procedure and potentially degrading performance.

A well-known alternative formulation of \glspl{de} is the weak formulation, in which the \gls{de} is expressed as an integral of the product of the residual and an arbitrary test function. This approach has been explored as a basis for loss functions in classical \gls{pi} architectures and offers several complementary advantages~\cite{kharazmi2019variational, wang2025wf}. The weak formulation is inherently global, naturally propagating boundary information across the domain. Furthermore, it enables the use of techniques such as integration by parts to redistribute derivatives and better encode boundary conditions, thereby discouraging trivial solutions and promoting generalization beyond training points. However, the weak formulation alone lacks the pointwise enforcement that collocation provides, potentially allowing localized errors to persist.

In this paper, we introduce the weak formulation as a loss function for quantum \glspl{pinn} and demonstrate that it addresses key limitations of collocation-based approaches. Recognizing that the two formulations offer complementary strengths—collocation provides strong pointwise enforcement, while the weak form ensures global consistency and boundary propagation — we investigate systematically the performance of a hybrid loss function that combines both terms. 
We validate the advantage of this combined approach by solving various \glspl{de} with domain decomposition, demonstrating improved convergence and reduced susceptibility to trivial solutions.

\section{Methodology}

We now detail our approach to solving \gls{de} problems in a variational manner, by including both a discrete set of collocation terms, and a global weak form regularization. 
First, we summarise the weak formulation principle for \glspl{de}. 
Then, we recall the fundamentals of physics-informed approaches in both classical and quantum architectures, referring mostly to the original publications for a complete description.  
Finally, we introduce the weak form in the procedure, showing how to leverage upon integration by parts to embed it into a single loss function, and ultimately how to benefit from it in an improved training workflow.

\paragraph{Formulations of a Differential Equation.}
We start considering the standard residual form of a \gls{de}, e.g. for a one-dimensional domain:
\begin{align}
    \mathrm{\gls{de}}(x, f, df/dx, ...) = 0 ~ \forall ~ x ~ \in ~\Omega.
    \label{eq:residual_form}
\end{align}
An alternative way to render Eq.~\ref{eq:residual_form} is via its weak form. 
we multiply the equation by another function $v(x)$, the \textit{test function}, to attain:
\begin{align}
    v(x) \mathrm{\gls{de}}(x, f, df/dx, ...) = 0 ~ \forall ~ v(x) ~ \forall ~ x ~ \in ~\Omega.
\end{align}
The latter condition can be replaced by integrating over the domain $\Omega$: 
\begin{align}
    \int_\Omega v(x) \mathrm{\gls{de}}(x, f, df/dx, ...) dx = 0 ~ \forall ~ v(x),
    \label{eq:weak_form}
\end{align}
which is the weak formulation of the original \gls{de} in Eq.~\ref{eq:residual_form}. 

\paragraph{(Quantum) physics-informed methods.}
\glspl{pinn} introduced the idea of adopting a trainable model to generate trial solutions $f(\mathbf{x})$ of a \gls{de}, tuned to minimize a set of constraints informed by the differential equation form, boundary and initial conditions~\cite{cai2021physics, cuomo2022scientific}. 
These constraints are combined together in a \textit{loss function}, which can then be minimised for the model to represent a faithful surrogate of the targeted solution, say $\bar{f}(\mathbf{x})$.
Therefore, the main components in the workflow of these algorithms can be identified with: (1) the chosen model, (2) the loss function and (3) the training procedure.

\paragraph{Trainable models}

There are many different, typically heuristic, choices to construct a trainable model in physics-informed scenarios. 
In the classical case, \glspl{mlp} can be replaced by e.g. convolutional \glspl{nn} or \gls{lstm} recurrent \glspl{nn}, and various constrained architectures among others. 
Also, different activation functions can be invoked, which can strongly affect the final performance of the model \cite{zhongkai2024pinnacle}. 
Replacing such classical architectures with a \gls{qnn} ~\cite{} - led to the development of the quantum counterparts of : \gls{dqc} \cite{kyriienko2021solving}. 
A \gls{qnn} consists of a quantum circuit where (some of) the gates feature trainable variational parameters, and others encode the problem features. Differentiation can be attained via specific parameter-shift rules, thus satisfying the main requirements of a \gls{pi} trainable model. 
We provide a more formal yet succinct description of the \gls{qnn} architecture in SI~\ref{app:models}.
Likewise for \glspl{qnn}, one can adopt various choices involving either the feature map~\cite{jaderberg2024let}, the ansatz construction \cite{setty2025selfqpinn} or the cost operator choice~\cite{kyriienko2021solving, paine2025cost}.

In this work, we use alternatively as the trainable models \glspl{mlp} or \glspl{qnn}, adopting a single, consistent standard architecture.
This is in line with the scope of the work, i.e. benchmarking the introduction of a global regularisation term, without attempting a simultaneous systematic benchmark of various possible design choices, which may be the subject of further analysis.

\paragraph{Loss function}
\label{sec:loss_function}
The loss function encodes the restraints associated with the \gls{de} - such that when minimised - the trial function resulting as the model output will approximate the solution to the given problem. 
This can be achieved in many ways, but the use of a discrete set of test points is the most common among general-purpose strategies. 
That is, known a restraint on the solution $f(\mathbf{x})$ formally implied by the residual form Eq.~\ref{eq:residual_form}, one can write (generalising to a higher dimensionality of the problem):
\begin{align}
    \mathcal{L}_{\text{DE}}(\theta) = \sum_j L(\text{DE}(\mathbf{x}_j, f_\theta(\mathbf{x}_j))),
    \label{eq:loss_residual}
\end{align}
where $\{ \mathbf{x}_j \}$ in the domain $\Omega$ are called the \textit{collocation points} and $L(\cdot)$ is a distance measure - such as the mean square error (MSE) or mean absolute error (MAE). 
A similar approach can be used to satisfy boundary conditions or initial values, that we label together as $\text{IBV}(\mathbf{x}, f(\mathbf{x})) = 0\; \forall x \in \delta\Omega$, by introducing additional loss terms
\begin{align}
    \mathcal{L}_{\text{IBV}}(\theta) = \sum_j L(\text{IBV}(\mathbf{x}^{\text{IBV}}_j, f_\theta(\mathbf{x}^{\text{IBV}}_j))),
    \label{eq:boundary_loss}
\end{align}
where $\mathbf{x}^{\text{IBV}}_j$ is a set of collocation points along the boundary of interest $\delta\Omega$. 

It must be noted that some variational terms in the losses $\mathcal{L}_{\text{DE}}, \mathcal{L}_{\text{IBV}}$ as above can be alternatively replaced by enforcing the corresponding constraints by design. This amounts to effectively bias the model, and can often reduce the computational workload especially in high-dimensional applications \cite{lu2022modlanets, ghosh2023harmonic, ghosh2025neural}.

\paragraph{The Weak Form as a regularization term} 
The weak formulation in Eq.~\ref{eq:weak_form} can be used as the requirement for a loss term, instead of the residue. This was considered classically for \glspl{pinn}~\cite{kharazmi2019variational, chaumet2022improving, wang2025wf}. Similar to how we choose a set of training points for the collocation based approach, we consider a chosen set of trial functions $\{v_j(x)\}_j$. A loss function can then be formed as 
\begin{align}
    \mathcal{L}_{\mathrm{WF}}(\theta) = \sum_j L \left( \int_\Omega v_j(x) \mathrm{\gls{de}}(x, f_\theta, df_\theta/dx, ...) dx \right),
    \label{eq:loss_weak}
\end{align}
equivalently to Eq.~\ref{eq:loss_residual}.
It is further possible to adapt the initial form Eq.~\ref{eq:loss_weak} with any technique to rewrite/adapt integrals, without affecting its validity as a loss term. 
One technique is \gls{ibp}:
\begin{align}
    \int_a^b v u' dx = \left[ v u\right]_a^b - \int_a^b v' u dx.
\end{align}
Due to the linearity of the integral operator, this can be performed on separate terms $u$ of the \gls{de} and allows for rearrangement of orders of derivatives between the trial function $f_\theta$ and test functions $v_j(x)$. Generally, derivatives for $f_\theta$ are more expensive to compute than for $v_j(x)$, therefore this procedure can ease the overall computational complexity. 
Additionally, this method greatly reduces the likelihood of $f(\mathbf{x})$ converging to trivial solutions that satisfy the \gls{de} over most of the domain, but for its boundaries. This is because by performing \gls{ibp}, the boundary terms get effectively included in satisfying the \gls{de}, altering the loss landscape and reducing the ``attractiveness'' of the trivial solution local minima. 
It is also interesting to note that collocation points can be considered a specific case of the weak form where the test functions are delta function $v_j(x) = \delta(x-x_j)$. However, because of the properties of these functions - discontinuous and local - the specific benefits mentioned when discussing the weak formulation do not hold.

\paragraph{Combining loss contributions.}

The residual loss function based on a point collocation strategy is typically formed of a summation of loss terms from each differential restraint in Eq.~\ref{eq:loss_residual} and Eq.~\ref{eq:boundary_loss} for the encoded initial/boundary values:
\begin{align}
    \mathcal{L}_{\text{RES}}(\theta) = \sum_j \alpha_j \mathcal{L}_{\text{DE}_j}(\theta) + \sum_j \beta_j \mathcal{L}_{\text{IBV}_j}(\theta)
\end{align}
where coefficients $\alpha$  and $\beta$ change the emphasis of the terms and collocation points, and can be chosen statically or be variational themselves. 
We have also discussed an alternative term inspired by the weak formulation $\mathcal{L}_{\text{WF}}$.

However, loss contributions can be combined by summing them together as:  
\begin{align}
    \mathcal{L}(\theta) = \gamma_{\text{RES}} \mathcal{L}_{\text{RES}}(\theta)  +
    \gamma_{\text{WF}} \mathcal{L}_{\text{WF}}(\theta)
    \label{eq:combination}
\end{align}

This alters the training behaviour and we will investigate the difference between considering one or another or both combined for specific examples. 
Additional regularisation terms $\mathcal{L}_{\text{REG}}$ can also be included - these terms tend to either encourage or discourage behaviour, generally to either improve generalisation or avoid local minima.

\paragraph{Domain Decomposition} 
We now examine a possible technique to use within \gls{pi} algorithms - domain decomposition \cite{smith1997domain, kharazmi2021hp}. 
This is when the global \gls{de} domain is split into separate subdomains, which are considered separately. A different model can then be assigned to each subdomain. This can greatly increase expressivity - especially useful if strongly different behaviour is expected in different regions of the domain. 

However, it does need to be ensured that the overall function is continuous over all sub domains and that the boundary/initial conditions propagates through to non-adjacent subdomains. 
In collocation based strategies, this is usually achieved with an additional loss term minimising the difference between adjacent models at the boundary such as
\begin{align}
    \label{eq:domdecomp_lossterm}
    \mathcal{L}_\mathrm{SBC}(\theta) = \sum_{j,k} L(f_\theta^{j+}(x^j_k) - f_\theta^{j-}(x^j_k)).
\end{align}
$j$ sums over all subdomain boundary's and $f_\theta^\pm$ denotes the trial functions encoding the subdomains bordering the $j^{\mathrm{th}}$ boundary. 

Crucially, though, a global loss term - as the one introduced by $\mathcal{L}_\mathrm{WF}$ - can automatically promote continuity as a training goal for the overall loss function. 
When operating with domain decomposition, this can thus be another possible benefit from a weak formulation: ``bridging'' the subdomains. 
It must be noted  though how simply introducing $\mathcal{L}_\mathrm{WF}$ in any form, this approach might not necessarily be enough by itself - we want both continuity in $f_\theta(\mathbf{x})$ across subdomain boundaries, but also the information on IBV to propagate into subdomains which aren't adjacent to such boundaries. 
However, the latter can be ensured by making sure that $f_\theta(\mathbf{x})$ appears in the appropriate integrals, e.g. via the aforementioned \gls{ibp}.

\paragraph{Training and its cost.}

The trial function $f_\theta(\mathbf{x})$ can then be trained, to minimize the chosen loss function. 
At first, the variational parameters are initialised to some values - at random or following an initialisation strategy. 
Then, an optimiser routine is chosen, one popular choice being ADAM \cite{kingma2014adam}, which across a training loop, suggests new parameters given the current evaluation of the loss and any derivatives relevant to the chosen optimizer. 
This repeats until an end condition is reached, either in terms of allocated steps or a loss variation within a predetermined convergence threshold. The resulting parameters then define a proposed solution.

A relevant question is whether the introduction of $\mathcal{L}_{\text{WF}}$, exclusively or not, bears any additional cost in quantum resources. 
The reasoning can be split in the two main contributions: the number of points being evaluated, and the order of derivation.
Concerning the first, if the numerical method for integration uses the same $\mathbf{x}$ values as the collocation points considered, then no additional cost ensues. 
In terms of order of derivation, adopting \gls{ibp} can result in additional terms being included, but they will always be of lower order than the highest order derivative in the \gls{de}, thus minimising the impact also in this regard.

\begin{figure}
        \centering
    \includegraphics[width=.7\linewidth]{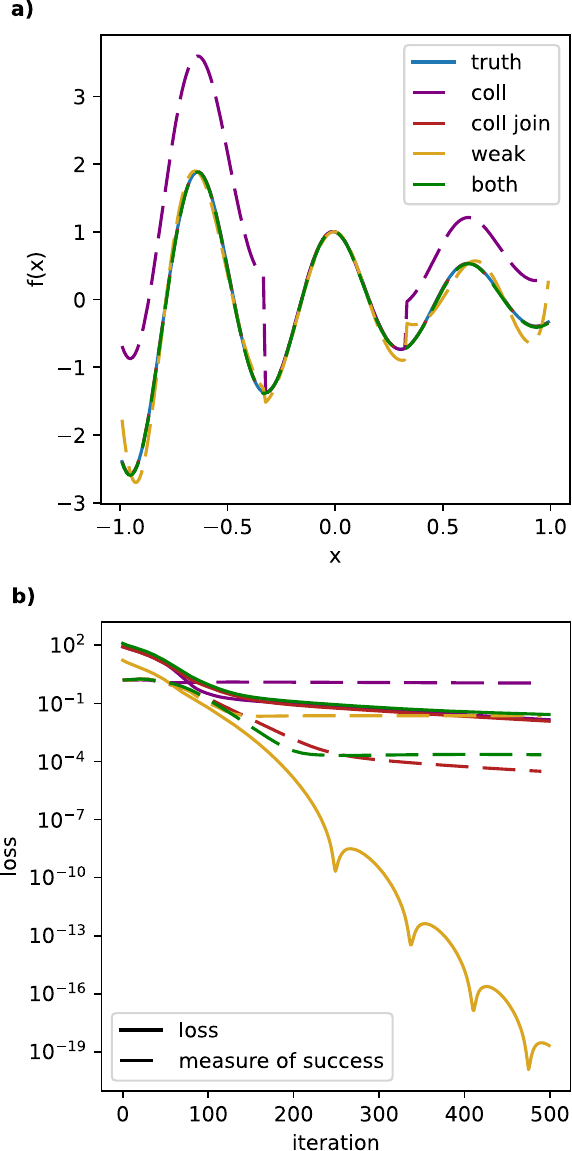}
    \caption{Attained solution (a) and training (b) of Eq.~\ref{eq:damped_osc}, comparing the baseline analytical solution (\emph{truth}) against the converged solutions calculated adopting as the relevant loss term(s) to train against: 
    (\emph{coll}) only the collocation-based Eq.~\ref{eq:loss_residual},  (\emph{coll join}) the latter, inclusive of $\mathcal{L}_{\mathrm{SBC}}$ from  Eq.~\ref{eq:domdecomp_lossterm}, 
    (\emph{weak}) the weak-formulation alone or
    (\emph{both}) a combination of both weak and collocation contributions, as seen in Eq.~\ref{eq:combination}. 
    The evolution of the latter is shown in (b) as a solid line, against training epoch and along with (as dashed line) the measure-of-success metric as defined in the main text.
    }
    \label{fig:damped_osc}
\end{figure}

\section{Results}

We now present a variety of examples, to compare the results of using collocation points, weak formulation and a combination of both as a strategy in constructing the total loss function $\mathcal{L}(\theta)$. 
All examples presented here are classically simulated with noise-free, state-vector simulation using \texttt{qadence}~\cite{seitz2025qadence}. 
We adopt MSE as the chosen distance function.
Details on the integration procedure are reported in SI~\ref{app:integration}.
When adopting domain decomposition, the same form of trainable (quantum) model is used for each subdomain, but with independently trained variational parameters. The optimiser used is ADAM, with learning rate $0.2$. 
As measure-of-success, we adopt the MSE of the solution inferred from each attempted trainable model, against the known analytical solution, over the same grid chosen for the training points.

\subsection{Damped Oscillator} \label{sec:damped_osc}

The first example is a first-order 1D linear \gls{de}:
\begin{align}
\label{eq:damped_osc}
    \frac{df}{dx} = & \kappa \exp(-\kappa x) \cos(\lambda x) + \lambda \exp(-\kappa x) \sin(\lambda x) \nonumber
    \\  f(0) = & 1,
\end{align}
with $\kappa = 1, \lambda = 10$. Its solution  $f(x) = \exp(-\kappa x) \cos(\lambda x)$ represents a damped oscillator. 

The domain is decomposed into $(-1, -0.33], [-0.33, 0.33], [0.33, 1)$. We consider 30 points uniformly distributed within each subdomain for a total of 90 points. These are the points for the collocation method, and form the numerical integration grid for the weak formulation. 

We solve this equation adopting domain decomposition, and a \gls{qnn} on a register of 5 qubits, tower Chebyshev feature map, depth-4 \gls{hea} and a classically post-processed observable 
\begin{equation}
    C =a \sum_{j=0}^{N-1} Z_j + b I,
    \label{eq:cost_function}
\end{equation}
where $Z_j$ is the magnetization of the $j$-th qubit, $a$ and $b$ are variational parameters controlling scale and shift respectively. 

For the weak formulation we consider $11$ sinusoidal test functions, with the IBV included in $\mathcal{L}_{\text{WF}}(\theta)$ as we use \gls{ibp}, as reported in more detail in SI~\ref{app:damped_oscillator}.
For the collocation method, we adopt $\mathcal{L}_{\text{DE}}$ on the chosen grid initial value considered.

The results of this training are shown in Fig. \ref{fig:damped_osc}. Collocation points only fits the \gls{de} well, but does not propagate the initial condition between the different subdomains. Weak from only does not have enough information from the relatively limited number of test functions to solve the \gls{de} perfectly but does capture the rough shape. Combining the two together results in a well trained function, as emphasised by the much better measure-of-success.

\begin{figure}
        \centering
    \includegraphics[width=.7\linewidth]{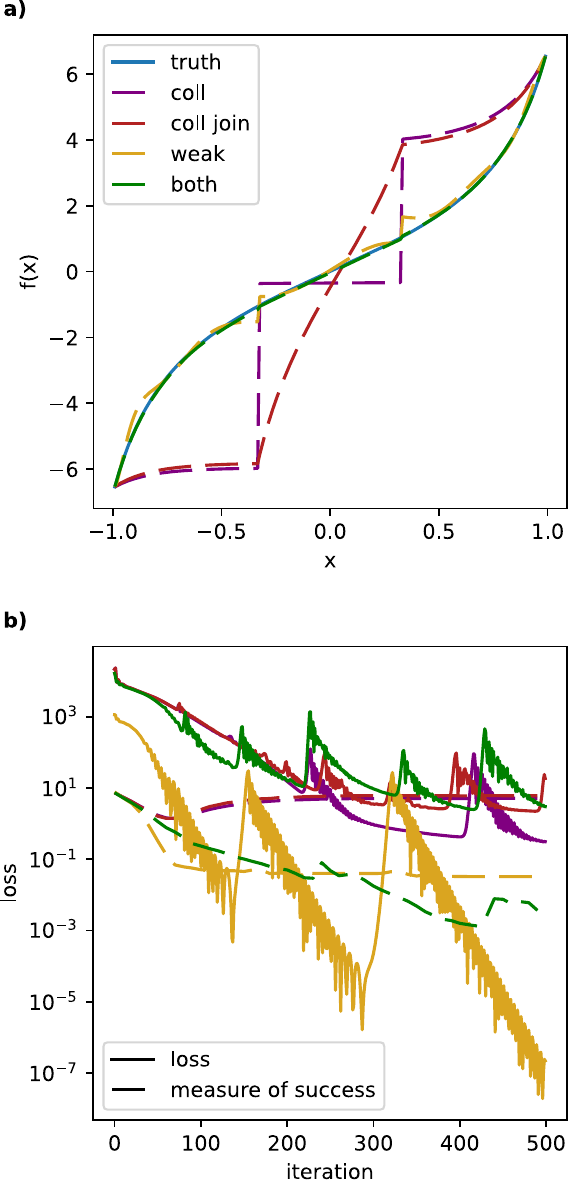}
    \caption{Attained solution (a) and training (b) of stationary Burgers Eq.\eqref{eq:stnry_burgers}.
    In (a) we report the baseline analytical solution (\emph{truth}) as a solid line, against the converged solutions adopting various strategies as dashed lines, labelled according to the caption of Fig.~\ref{fig:damped_osc}, to provide the relevant loss function for the training. 
    The evolution of the latter is shown in (b) as a solid line, against training epoch and along with (as dashed line) the measure-of-success metric as defined in the main text.
    }
    \label{fig:stnry_burgers}
\end{figure}

\subsection{Stationary Burgers Equation} \label{sec:stnry_burgers}

We now move on to consider the stationary Burgers equation:
\begin{align}
\label{eq:stnry_burgers}
    f \frac{df}{dx} - \nu \frac{d^2f}{dx^2} =0, ~~ f(\pm 1) = f_{\pm 1},
\end{align}
with $\nu = 1$. This is a second-order, nonlinear \gls{de} with known solution $f(x) = \sqrt{2 \nu a} \tan \left( \sqrt{\frac{a}{2 \nu}} (x+b) \right)$ with $\kappa$ and $\lambda$ two parameters determined by choice of initial/boundary condition. We choose $f_{\pm 1}$ such that $\kappa = 3$ and $\lambda=0$.

We choose the same experimental setup as in the previous example (including the test functions), with the only exception of increasing the variational ansatz to depth-8. 
The same test functions are used, leading though to a different $\mathcal{L}_{\text{WF}}(\theta)$, reported in SI~\ref{app:burgers}.

The result of this case upon training is shown in Fig.~\ref{fig:stnry_burgers}. This case highlights a typical situation where a collocation-only strategy converges to trivial, flat solutions which solve the \gls{de} without satisfying properly the boundary conditions. On the contrary, a pure weak formulation captures the overall behaviour of the solution within the domain, but given the frugal setup established by the chosen test functions in Eq.~\ref{eq:funcs_damped_osc} lacks enough information by itself to produce an accurate solution. Using both loss contributions again enhances the performances of both approaches when taken separately, producing a quantitatively good fit with the analytical solution.

\subsection{Linear 2D Example} \label{sec:simple_2D}

We now test the performance of our benchmark methods when increasing the dimensionality of the problem, and specifically using a linear 2D example:
\begin{align}
\label{eq:simple_2D}
    &\frac{df}{dx} + \frac{df}{dy} - 2(x+y) = 0, \nonumber \\ 
    & f(0,y) = y^2, ~~ f(x,0) =  x^2, 
\end{align}
within the domain defined by $x,y \in [0,1]$. This problem has known solution $f(x,y) = x^2 + y^2$.

The domain is split into four equal subdomains (i.e. adopting $x=0.5$ and $y=0.5$ as boundaries). 
A training grid is formed of 20 points uniformly spread across both $x$ and $y$, totalling $400$ points. 
For the quantum model a tower Fourier map encodes $x$, followed by a depth 1 \gls{hea}, followed by $y$ encoded with a tower Fourier map, followed by a depth 8 \gls{hea}. The observable is the same as in Eq.~\ref{eq:cost_function}.

\begin{figure*}
    \centering
    \includegraphics[width=0.52\linewidth]{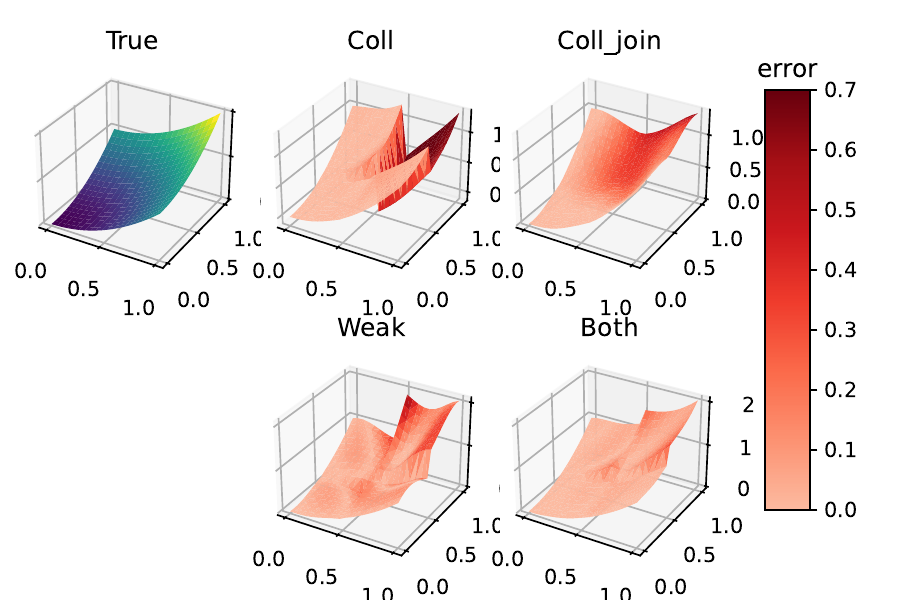}
    \includegraphics[width=0.47\linewidth]{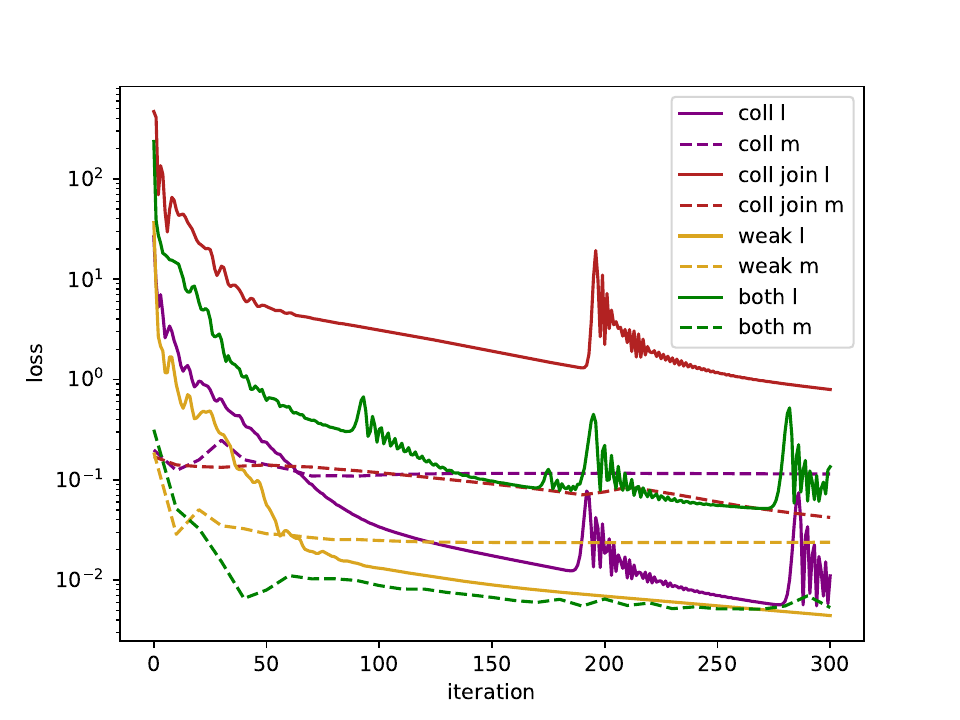}
    \caption{Attained solution (left) and training (right) of linear 2D Eq.\eqref{eq:simple_2D}.
    In the 3D plots we report the baseline analytical solution (\emph{True}) against the converged solutions adopting various strategies, labelled according to the caption of Fig.~\ref{fig:damped_osc}, to provide the relevant loss function for the training. 
    The evolution of the latter is shown in (b) as a solid line (\emph{l}), against training epoch and along with (as dashed line) the measure-of-success metric (\emph{m}) as defined in the main text.
    The 3D plots displaying the trained solutions colour-code the difference against the baseline solution (colour-coded against the value of $f(x,y)$ at each point. 
    }
    \label{fig:simple_2D}
\end{figure*}

Results are reported in Fig.~\ref{fig:simple_2D}, again displaying the advantageous combination of collocation and weak contributions to the loss, particularly in order to use successfully a domain decomposition strategy.

\subsection{Laplace} 
\label{sec:laplace}
 
 We finally consider the Laplace equation
 \begin{align}
 \label{eq:laplace}
     \frac{d^2f}{dx^2} + \frac{d^2f}{dy^2} &= 0 \nonumber
     \\f(0,y) = f(x, 0) = f(x, 1) &= 0 
     \nonumber
     \\ f(0,y) &= \sin(\pi y),
 \end{align}
 within a square box spanning $x$ and $y$ both between 0 and 1. This has known solution $f(x,y) = \sinh \left(\pi (x-1)\right)/ \sinh(- \pi) \sin(\pi y)$, with the solution $f(x,y) = 0$ offering a trivial solution satisfying all but the last boundary conditions in Eq.~\ref{eq:laplace}.

To solve this problem, we split the domain in the same way as in Sect.~\ref{sec:simple_2D}, with a training grid of 21 points in both $x$ and $y$, totalling $441$ points. 
In order to keep the computational cost comparable (as it now involves second order derivatives), we reduce the quantum model to a narrower register of $4$ qubits. 
$x$ and $y$ are uploaded with tower Fourier feature map, separated by a depth 1 \gls{hea}. They are re-uploaded a total of twice each, before a depth-6 final variational \gls{hea} layer. 
Given the increased complexity of the problem, we move to consider $144$ test function in order to formulate an appropriate $\mathcal{L}_{\text{WF}}$.

\begin{figure*}
    \centering
    \includegraphics[width=0.52\linewidth]{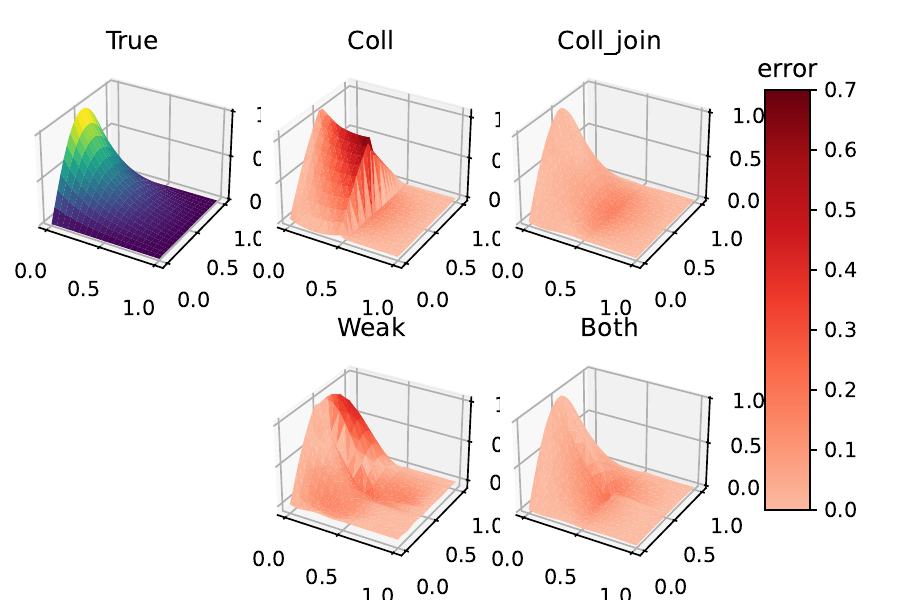}
    \includegraphics[width=0.47\linewidth]{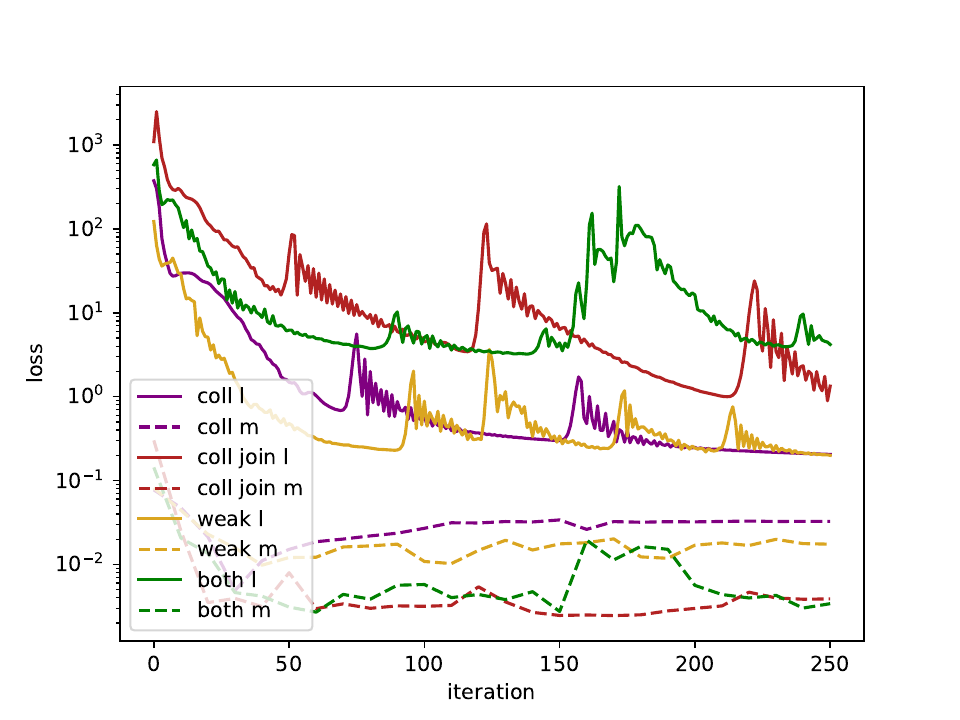}
    \caption{Attained solution (left) and training (right) of Laplace Eq.\eqref{eq:laplace}.
    In the 3D plots we report the baseline analytical solution (\emph{True}) against the converged solutions adopting various strategies, labelled according to the caption of Fig.~\ref{fig:damped_osc}, to provide the relevant loss function for the training. 
    The evolution of the latter is shown in (b) as a solid line (\emph{l}), against training epoch and along with (as dashed line) the measure-of-success metric (\emph{m}) as defined in the main text.
    The 3D plots displaying the trained solutions colour-code the difference against the baseline solution (colour-coded against the value of $f(x,y)$ at each point. 
    }
    \label{fig:laplace}
\end{figure*}

Results are shown in Fig. \ref{fig:laplace}, displaying the strong tendency of a pure collocation strategy to converge on trivial solutions, particularly away from the boundaries, and failing to reconstruct even qualitatively the correct shape of the solution. 
On the contrary, the weak form captures well the overall behaviour as illustrated in the previous example as well, but quantitatively is on-par with the collocation strategy and benefits from an additional loss contribution based upon collocation points to improve beyond.

\section{Discussion}

We have introduced and discussed the potential benefits of introducing a weak form \emph{along} with, rather than \emph{instead} of, collocation points in the loss function to train \gls{pi} models to represent correctly the solutions of differential problems. 
By systematically comparing their behaviour when solving a set of (non)linear 1D\&2D \glspl{de}, we could observe the different behaviours of each, and confirm numerically how the two approaches can compensate each other's shortcomings. 
This can be particularly attractive when adopting models of limited expressivity, as it can be the case in proof-of-concept experiments running on \gls{nisq} devices~\cite{philip2025experimental}. This motivated us to adopt as the benchmark architecture parsimonious quantum models involving registers spanning very few (4-5) qubits, and shallow ans\"atze.

The weak formulation is strong at propagating boundary information and hence regularising the trial solutions against that information. 
This penalises the collapse to trivial solutions and automatically provides information to join subdomains, particularly useful when using domain decomposition strategies for expressivity reasons. 
On the other side, including information at specific collocation points pinches the trial solutions to satisfy specific constraints, e.g. flexibly highlight subdomains where the accuracy of the solution is most stringent. 
When hybridising these formulations, we could observe the impact of different loss contributions when combined into a generalised loss function. 

Overall, we report significant advantages in the training and convergence of this hybrid methodology, focusing in particular on scenarios exhibiting trivial solutions as reasonable candidates to solve the target equations, as well as a domain decomposed in different training subdomains. 
Domain decomposition also partially acts as a generalisation test, as with little information being ported across the boundaries of the subdomains, the models are prone to overfit the training loss within the subdomain, performing poorly otherwise. 

These conditions are known to present challenges when attempting to capture solutions with pure collocation-informed strategies, but the introduction of relatively modest contributions from a weak-form regularization can much better inform the model towards the expected solution.
Interestingly, we also observed how the gap among the two methodologies is reduced when the trivial solution can satisfy the constraints very well on-average, as it was the case for the 2D Laplace equation, i.e. collocation-based losses and overall even a measure-of-success like MSE can highlight a good fit performance, despite the trial solution failing at capturing some essential features in selected points. In such cases, global contributions to loss terms can thus play an important role in recovering fully the solution behaviour. 

We briefly discussed the computational impact of entertaining weak-formulated loss terms, arguing how a proper selection of the grid does not alter significantly the overall cost of the training procedure when compared with equivalent collocation-only training strategies, especially when compared with the improvement in the quality of the results within very few training epochs. 
As for the scaling of test functions required - as the chosen problem scales in dimensionality and order of derivation - we did not perform independent testing as we refer to the abundant literature studying weak formulations for general-purpose \gls{de} solving - e.g. \cite{messenger2024weak, demkowicz2011class}. 

In conclusion, we believe our findings support a systematic introduction of weak-formulated contributions when training physics-informed models.

\newpage

\appendix
\section*{Supplementary Information}

\subsection{Integration techniques}
\label{app:integration}
To evaluate the loss function using the weak formulation, integrals need to be evaluated. Generally, we will need to evaluate this numerically -  analytic solutions either being unknown or too expensive to evaluate whilst training. Any appropriate numerical integration method for the problem is valid.

One possible approach is utilising the trapezium rule
\begin{align}
    \int_\Omega f(x) dx \approx \sum_{j=0}^{M-1} \frac{x_{j+1}-x_j}{2}(f(x_j) + f(x_{j+1})),
\end{align}
over a set of evaluation points $\{x_j\}_j$ chosen from 
the domain - similar to the choice of training points when considering the collocation point form of loss.

Another possible method is Monte-Carlo integration
\begin{align}
    \int_\Omega f(x) dx \approx V/M \sum_{j=0}^{M-1} f(x_j),
\end{align}
where $V = \int_\Omega dx$. This method generally requires large M.

\subsection{Differentiable Quantum Circuits}
\label{app:models}

More formally, the \gls{qnn} operation can be described in blocks: starting from an initial state $| \emptyset \rangle$, some blocks of gates - described as the \gls{fm} unitary $\hat{\mathcal{U}}(\mathbf{x})$ - are dedicated to encode the problem's variables $\mathbf{x}$, with other blocks parametrising instead a variational ansatz $\hat{\mathcal{V}}_\theta$, via a set of continuous angles $\boldsymbol{\theta}$. 
The information encoded in the evolved state is then accessed via a measurement operator $\hat{\mathcal{C}}$. 
The output of this architecture can then be synthetically described with the expectation value:
\begin{align}
    \mathrm{QNN}_{\theta} (\mathbf{x}) = \langle \emptyset | \hat{\mathcal{U}}^\dag(\mathbf{x}) \hat{\mathcal{V}}^\dag_\theta \hat{\mathcal{C}} \hat{\mathcal{V}}_\theta \hat{\mathcal{U}}(\mathbf{x}) | \emptyset \rangle
\end{align}
Crucially, derivatives against such parameters can be evaluated analytically by using the parameter shift rule \cite{schuld2019evaluating} or some generalisation thereof \cite{kyriienko2021generalized, agpsr}.
The output $\mathrm{QNN}_{\theta} (\mathbf{x})$ can be additionally post-processed via inexpensive classical operations (e.g. shifting or rescaling) to represent a trial function $f_\theta(\mathbf{x})$.

\subsection{Experimental details}

\paragraph{Damped Oscillator}
\label{app:damped_oscillator}

The test functions for the weak formulation are:
\begin{align}
    v_j(x) = \begin{cases} \cos(\pi j x) ~~ j<0 \\
                            1 ~~ j = 0 \\
                            \sin(\pi j x) ~~ j>0,
    \end{cases}
    \label{eq:funcs_damped_osc}
\end{align} 
considering $j$ integer labels in $[-5, 5]$

The $\mathcal{L}_{\text{WF}}$ term for each test function is then
\begin{align}
    \mathcal{L}_{v_j}(\theta) = \big(&[f_\theta(x) v_j(x)]_{-1}^1 - \int_{-1}^1 f_\theta(x) v_j'(x) dx \nonumber \\ 
    &+ \int_{-1}^1 v_j(x) (\kappa \exp(-\kappa x) \cos(\lambda x) \nonumber \\ 
    &+ \lambda \exp(-\kappa x) \sin(\lambda x) ) dx\big)^2,
\end{align}

The optimiser used is ADAM with learning rate $0.2$.

\paragraph{Burgers equation}
\label{app:burgers}

In this case, integration by parts of Eq.~\ref{eq:stnry_burgers} adopting the same test functions as in Eq.~\ref{eq:funcs_damped_osc} results in a weak formulation loss consisting of terms
\begin{align}
    \mathcal{L}&_{v_j}(\theta) = \big([\nu f_\theta(x) v_j'(x)]_{-1}^{1} - [\nu f_\theta'(x) v_j(x)]_{-1}^1 \nonumber \\ &- \nu \int_{-1}^1 f_\theta(x) v_j''(x) dx + \int_{-1}^1 f_\theta(x) f_\theta'(x) v_j(x) \big)^2.
\end{align}

\paragraph{Linear 2D equation} 

For the test functions we here considered of the form
\begin{align}
    v_{j,k}(x, y) = \cos(\pi j x) \sin(\pi k y).
\end{align}
In particular, we consider those with labels of $j$ and $k$ iterating over the integers between $1$ and $3$. The term for each test function (after integration by parts) for the weak formulation is then
\begin{align}
    \mathcal{L}_{v_{j,k}}(\theta) = \Big(&\int_0^1 [f_\theta(x,y) v_{j,k}(x,y)]_0^1 dy \nonumber\\
    &-\int_0^1 \int_0^1 f_\theta(x,y) \frac{d v_{j,k}(x,y)}{dx} dx dy \nonumber \\
    &+ \int_0^1 [f_\theta(x,y) v_{j,k}(x,y)]_0^1 dx \nonumber \\
    &-\int_0^1 \int_0^1 f_\theta(x,y) \frac{d v_{j,k}(x,y)}{dy} dx dy \nonumber \\ 
    &-2 \int_0^1 \int_0^1 (x+y) v_{j,k}(x,y) dx dy\Big)^2.
 \end{align}
 The integrals are calculated with 1D and 2D trapezium rules using the grid discussed in the main text.

 \paragraph{Laplace equation}
 \label{app:laplace}

 In this case, for test functions we consider 
 \begin{align}
    v_{j,k}(x, y) = \cos(\pi j x) \sin(\pi k y).
\end{align}
In particular, we consider $144$ such functions with $j$ and $k$ chosen randomly between 0.1 and 10. The weak loss term for each test function (after integration by parts) is then
\begin{align}
    \mathcal{L}_{v_{j,k}}(\theta) = \Big( \int_0^1 \int_0^1 f_\theta(x,y) \left(\frac{d^2 v_{j,k}}{dx^2} + \frac{d^2 v_{j,k}}{dy^2}\right) dx dy \nonumber \\
    + \int_0^1 \sin(\pi y) \frac{d v_{j,k}}{dx} dy - \int_0^1 \frac{df_\theta(0,y)}{dx} v_{j,k}(0,y) dy \nonumber \\
   +  \int_0^1 \frac{df_\theta(1,y)}{dx} v_{j,k}(1,y) dy + \int_0^1 \frac{df_\theta(x,1)}{dy} v_{j,k}(x,1) dx \nonumber \\
   - \int_0^1 \frac{df_\theta(x,0)}{dy} v_{j,k}(x,0) dx. 
\end{align}

\bibliography{bibliography}

\end{document}